\documentclass{aa}
\usepackage{txfonts}
\usepackage{graphicx}
\graphicspath{{./images/}}
\usepackage{hyperref}
\usepackage{natbib}
\bibpunct{(}{)}{;}{a}{}{,}
\usepackage{xcolor}
\usepackage{makecell}
\usepackage{tablefootnote}

\hypersetup{
    colorlinks=true,
    linkcolor=blue,
    citecolor=blue,
    filecolor=magenta,      
    urlcolor=blue,
}

\newcommand{\addone}[1]{#1}

\author{
    Simon Müller\inst{1}
    \and
    Jana Baron\inst{1}
    \and
    Ravit Helled\inst{1}
    \and
    Fran\c{c}ois Bouchy\inst{2}
    \and 
    Léna Parc\inst{2}
}

\authorrunning{Müller et al.}

\title{The mass-radius relation of exoplanets, revisited}

\institute{
    Department of Astrophysics, University of Zürich \\
    Winterthurerstrasse 190, 8057 Zürich, Switzerland 
    \and
    Observatoire de Gen{\`e}ve, Universit{\'e} de Gen{\`e}ve, 51 \\Chemin Pegasi, 1290 Versoix, Switzerland\\
    \email{simonandres.mueller@uzh.ch}
}

\date{Received 21 November 2023 / Accepted XX XX XXXX}

\abstract{
Determining the mass-radius ($M$-$R$) relation of exoplanets is important for exoplanet characterization. Here we present a re-analysis of the $M$-$R$ relations and their transitions using exoplanetary data from the PlanetS catalog which only includes planets with reliable mass and radius determinations. We find that "small planets" correspond to planets with masses of up to \addone{$\sim4.4 M_\oplus$ (within 17\%) where $R \propto M^{0.27}$}. Planets with masses between \addone{$\sim4.4$ and $127 M_\oplus$ (within 5\%)} can be viewed as "intermediate-mass" planets, where \addone{$R \propto M^{0.67}$}. Massive planets, or gas giant planets, are found to have masses beyond \addone{$127 M_\oplus$ with an $M$-$R$ relation of $R \propto M^{-0.06}$}. By analyzing the radius-density relation we also find that the transition between "small" to "intermediate-size" planets occurs at a planetary radius of \addone{$\sim1.6 R_\oplus$ (within 3\%)}. Our results are consistent with previous studies and provide an ideal fit for the currently-measured planetary population.
}

\begin{document}

\maketitle

\section{Introduction}\label{sec:introduction}

Since \addone{detecting} the first exoplanet around a sun-like star in 1995 \citep{mayor_jupiter-mass_1995} over 5000 exoplanets have been discovered revealing a large diversity in their physical properties. The field of exoplanets is blossoming: We are at a stage where we move from exoplanet detection to exoplanet characterization, with both of these fronts being extremely active and reaching a new level. \addone{For} exoplanet characterization, the two fundamental parameters are the planetary mass and radius. However, these two properties cannot be measured by the same method, and unfortunately in many cases only one of the two is available. To allow for a broader overview of exoplanets in a statistical sense, it is valuable to determine their mass-radius ($M$-$R$) relation and how this relation changes for different planetary types. \addone{Relating} a measured radius to a planet's mass can also help with radial-velocity follow-ups by estimating the expected radial velocity semi-amplitude of a transiting exoplanet. 

The $M$-$R$ relation depends on the planetary composition and therefore on the behavior of the different materials at planetary conditions  \cite[e.g.,][]{seager_mass-radius_2007, 2009AIPC.1094..102C, 2009ApJ...693..722G, 2012A&A...547A.112M,2014PNAS..11112622S, 2019AREPS..47..141J}. Theoretical $M$-$R$ relations can be inferred from interior models that rely on equations of state (EOS), which relate the density and pressure (and \addone{usually} also the temperature) of a given composition. For simplicity, often small terrestrial planets are assumed to have constant densities and are thus expected to behave as $R \propto M^{1/3}$ \cite[e.g.,][]{spiegel_structure_2014}. On the other hand, in massive giant planets that are hydrogen-helium (H-He) dominated in composition, the gravitational pressure is high enough for the materials to be pressure ionized and the electron degeneracy pressure becomes substantial \citep[e.g., ][]{2020NatRP...2..562H}. \addone{This causes the radius to decrease with increasing mass}, with $R \propto M^{-1/3}$. Of course, the planetary radius does not only depend on its mass but also other factors like stellar irradiation or the planetary age. 

\addone{Today} there are enough planets with mass and radius measurements to statistically infer an observed $M$-$R$ relation. The observed $M$-$R$ relations of exoplanets have been investigated in several studies and are often fitted by a broken power law. The breakpoints are \addone{particularly interesting} as they represent the transitions between different planetary types. For example, \citet{weiss_mass_2013} inferred a mass-radius-incident-flux relation. Based solely on the visual inspection of the $M$-$R$ and mass-density ($M - \rho$) diagrams, they found a transition point at $150 M_\oplus$. They inferred relations $R \propto M^{0.53} F^{-0.03}$ for $M < 150 M_\oplus$ and $R \propto M^{-0.04} F^{0.09}$ for $M > 150M_\oplus$, where $F$ is the instellation flux. \addone{They} suggested that for small planets the mass is the most important parameter for predicting the planetary radius, whereas for giant planets the incident flux is more important. Below $150 M_\oplus$ only \addone{35 planets were} available at the time, and the result for this region is less robust. 

A different approach was taken by \citet{hatzes_definition_2015}. In this study, the transition for giant planets \addone{was} explored. They used observations to infer the \addone{mass-density ($M$-$\rho$)} relation. A minimum in density at $\simeq 95 M_\oplus$ ($0.3 M_J$) and a maximum at $\simeq 1.9 \times 10^4 M_\oplus$ ($60 M_J$) were inferred, suggesting that these values correspond to the transition into gas giant planets. This study also suggested \addone{there} is no separation between brown dwarfs and giant planets as they display similar behavior. The mass-density relation of giant planets was determined to be $\rho = 0.78 M^{1.15}$.

\citet{chen_probabilistic_2016} presented an elaborate Markov Chain Monte Carlo method to analyze the $M$-$R$ relations of objects ranging from dwarf planets to stars. They inferred four distinct regions in their work: Terran worlds, Neptunian worlds, Jovian worlds, and stars. The corresponding transition points were fitted and identified as 2.04, 132, and 2.66 $\times 10^{4} M_\oplus$, where the latter two values correspond to $0.41 M_J$ and $0.08 M_\odot$. The $M$-$R$ relations were identified as: $R \propto M^{0.28}$ for Terran worlds, $R \propto M^{0.59}$ for Neptunian worlds, $R \propto M^{-0.04}$ for Jovian worlds and $R \propto M^{0.88}$ for stars. The results were obtained by a solely data-driven analysis rather than being derived from any physical assumptions. At the time of the study, only a few objects at $\sim 1 M_\oplus$ had been observed, and therefore the transition point between the Terran and Neptunian worlds relied on a small number of planets. 

\citet{bashi_two_2017} fitted the $M$-$R$ relations of two distinct regions using a total least squares approach, where the transition point was \addone{assumed to be} unknown. It was found that the transition occurs at $\sim 124 M_\oplus$ and $12.1 R_\oplus$. The data was best fitted by the relations: $R \propto M^{0.55}$ for small planets and $R \propto M^{0.01}$ for large planets. In this study, the transition point was also fitted instead of being imposed by some prior assumption. 

\citet{otegi_revisited_2020} presented an updated catalog of exoplanets, for which robust measurements of both radius and mass are available, based on the NASA Exoplanet Archive catalogue\footnote{\url{exoplanetarchive.ipac.caltech.edu}}. \addone{They focused on finding the transition} between rocky planets and those with a substantial gas envelope, and therefore only planets with masses up to $120 M_{\oplus}$ were considered. When displaying the planets in the $M$-$R$ plane, two distinct regions were identified: the rocky and the volatile-rich populations. Because they overlap in mass and radius, the populations were separated by the pure-water composition line to distinguish volatile-rich planets from terrestrial ones. For the \addone{two} distinct groups, the following ($M$-$R$) relations were inferred: $R = 1.03 M^{0.29}$ for the rocky population and $R = 0.70 M^{0.63}$ for the volatile-rich population.

\citet{edmondson_breaking_2023} showed that a discontinuous $M$-$R$ relation, as well as a temperature dependence for giant planets, results in a good fit to the $M$-$R$ measurements. Similar to \citet{otegi_revisited_2020} they separate between the rocky and the icy planets by a pure-ice EOS, while finding the transition of icy planets to gas giants at $115 M_\oplus$. For the rocky planets, they fit a relation of $R \propto M^{0.34}$ and for the icy planets $R \propto M^{0.55}$. When also considering the equilibrium temperature of the gas giants they find that $R \propto M^{0.00} T^{0.35}$. \addone{This suggests that} for the giant planets, the radius only depends on the temperature.

Recently, \citet{2023MNRAS.525.3469M} used a machine-learning approach to analyze the exoplanet population. By applying various clustering algorithms they identified the transition between "small" and giant planets at masses of $52.48 M_\oplus$ and sizes of $8.13 R_\oplus$. For the small planets, the $M$-$R$ relation was found to be $R \propto M^{0.50}$. They also showed that the radii of giant planets are positively correlated with the stellar mass. 

In this paper, we investigated the $M$-$R$ relations of exoplanets and their transitions using the PlanetS catalog\footnote{\url{dace.unige.ch}}. We used a solely statistical approach to determine the breakpoints in the relation used to define the different planetary regimes and determine the distinct dependencies. We also analyzed the mass-density and radius-density ($R$-$\rho$) relations and investigated their validity in separating different planetary types.

\section{Methods}\label{sec:methods}

In this work, we used the data from the PlanetS Catalog. An earlier version was presented in \citet{otegi_revisited_2020}; since then the catalog \addone{was} extended with additional discoveries and planets with masses up to $30 M_J$. There \addone{was also} an update on the planetary masses, and some planetary parameters \addone{were} reanalyzed. The catalog only includes planets with relative measurement uncertainties on the mass \addone{and radius smaller than 25 and 8\%}. Since the updated catalog contains many more planets, it is more reliable and allows for additional analyses. The data we used was downloaded in July 2023 and contains the mass and radius measurements of 688 exoplanets. Fig. \ref{fig:mass_radius_fit} shows how the planets are distributed in the $M$-$R$ plane.

Similar to previous work, our approach assumed a power-law relation between two planetary variables, for example, $M$ and $R$. The first step was to transform the variables into the log-log plane to use a linear regression method. We further assumed that there is an unknown number of break points in the linear relation between the two log-transformed variables, i.e., that a piece-wise linear function describes the data. 

\addone{For a single breakpoint, the two-segmented piece-wise linear relation can be parametrized as \citep[e.g.,][]{muggeo_estimating_2003}}:

\begin{equation}\label{eq:piecewise_linear_1bp}
    y = c + \alpha \xi + \beta(\xi - \psi) H(\xi -\psi) \, ,
\end{equation}

\addone{where $c$ is the intercept, $\alpha$ the slope of the first segment, $\beta$ the difference between the slopes of the two segments, $\psi$ the breakpoint and $\xi$ the independent variable. $H(x)$ is the Heaviside step function, defined as $H(x) = 1$ if $x \geq 0$ and $H(x) = 0$ otherwise.} 
\addone{For \textit{n} breakpoints, Eq. \ref{eq:piecewise_linear_1bp} can be generalized as:}

\begin{equation}\label{eq:piecewise_linear_nbp}
    y = c + \alpha_1 \xi + \sum^n_i \beta_i(\xi - \psi_i) H(\xi -\psi_i) \, .
\end{equation}

\addone{Because of the Heaviside step function Eqs. \ref{eq:piecewise_linear_1bp} and \ref{eq:piecewise_linear_nbp} are non-linear, and ordinary linear regression methods cannot fit the parameters. Furthermore, the data used in this work has significant uncertainties on the dependent and independent variables, which need to be considered.  Therefore, in order to determine the piece-wise linear regression fit, we used orthogonal distance regression (ODR) (as implemented in the Python package SciPy).}

\addone{We treated the number of breakpoints $n$ as an additional free parameter: To determine the number of breakpoints that yield the best fit to the data, we fit piece-wise linear functions with zero to four breakpoints. After fitting, we compared the models by calculating the Bayesian Information Criterion (BIC) in the form that is commonly used for linear regression:}

\begin{equation}\label{eq:bic}
    \textrm{BIC} = n \ln{\left(\frac{1}{n} \sum_i\left(x_i - \hat{x_i}\right)^2\right)} + k \ln{n} \, ,
\end{equation}

\addone{where $x_i$ are the data points, $\hat{x_i}$ the model predictions, and $n$, $k$ are the number of data points and model parameters.}

For the $M$-$R$ relation, $\xi \equiv \log(M[M_\oplus])$ and $y \equiv \log(R[R_\oplus])$, for the $M$-$\rho$ relation $xi \equiv \log(M[M_\oplus])$ and $y \equiv \log \rho$ [g/cm$^3$], and for the $R$-$\rho$ relation $xi \equiv \log(R[R_\oplus])$ and $\xi \equiv \log \rho$ [g/cm$^3$]. The planetary bulk density was calculated using $\rho = \frac{3 M}{4 \pi R^3}$.

\addone{Since all fitting variables required the data to be transformed, the uncertainties had to be propagated. For transforming the mass-radius measurements into their logarithms, the error propagation is simply $\sigma_\xi = \partial \left(\log x / \partial x\right) \sigma_x = \sigma_x / x \ln{10})$, with $x$ the measured mass or radius. When the measurement uncertainties were asymmetric, we used their arithmetic mean as $\sigma_x$.}

\addone{For vector-valued functions $f$ of potentially covariant parameters, the more general form of the error propagation has to be used \citep[e.g.,][]{doi:10.1021/jp003484u}:}

\begin{equation}
    \sigma^2_f = \mathbf{g}^T \, \mathbf{V} \, \mathbf{g} \, ,
    \label{eq:error_propagatiom}
\end{equation}

\addone{where $\mathbf{g}$ is the gradient of $f$ whose $i$-th element is $\partial f / \partial x_i$, and $\mathbf{V}$ is the covariance matrix. In our case, this applied to the uncertainties of the density and of the power-law parameters that were calculated from the piece-wise linear fit.}

\section{Results}\label{sec:results}

In this section, we first present our results for the $M$-$R$, $M$-$\rho$ and $R$-$\rho$ relations in Subsections \ref{sec:mass_radius_relation}, \ref{sec:mass_density_relation} and \ref{sec:radius_density_relation}. We then compare our results to previous studies in Subsection \ref{sec:comparison}.

\subsection{The mass-radius relation}\label{sec:mass_radius_relation}

\begin{figure}[h]
    \includegraphics[width=\columnwidth]{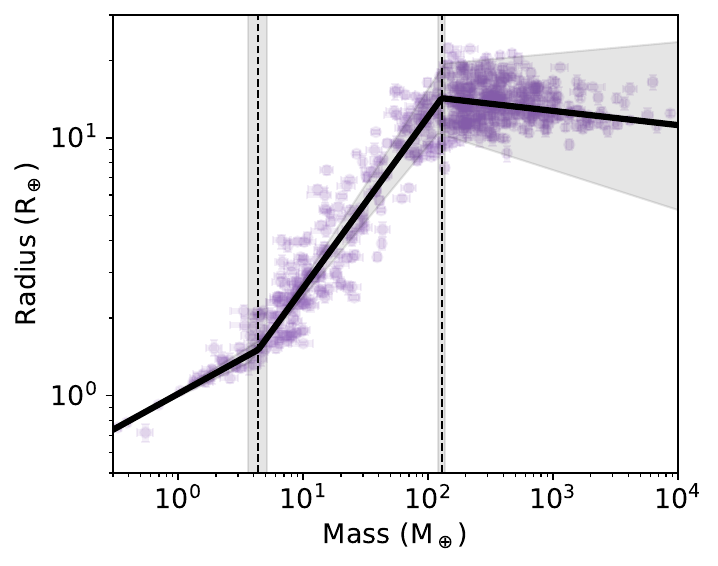}
    \caption{The data from the PlanetS Catalog is displayed in purple. The best-fit mass-radius relation (Eq. \ref{eq:mass_radius_relation}) is displayed as the solid black line and the shaded region shows the $1\sigma$-uncertainty. The breakpoints are shown as the dashed lines.}
    \label{fig:mass_radius_fit}
\end{figure}

\addone{By comparing the BIC of piece-wise linear models with zero to four breakpoints, we determined that two breakpoints provided the best fit to the $M$-$R$ distribution of the exoplanets from the PlanetS Catalog.} This led to the following $M$-$R$ relation:


\begin{equation}
    R = 
    \begin{cases}
        (1.02 \pm 0.03) \, M^{(0.27 \pm 0.04)} & M < (4.37 \pm 0.72) \\
        (0.56 \pm 0.03) \, M^{(0.67 \pm 0.05)} & (4.37 \pm 0.72) < M < (127 \pm 7) \\
        (18.6 \pm 6.7) \, M^{(-0.06 \pm 0.07)} & M > (127 \pm 7)
    \end{cases}
    \label{eq:mass_radius_relation}
\end{equation}

where $R$ and $M$ are in Earth units. The piece-wise linear fit with two breakpoints is shown together with the data in Fig. \ref{fig:mass_radius_fit}, and the fit parameters (see Eq. \ref{eq:piecewise_linear_nbp}) are listed in Tab. \ref{tab:results_mass_radius_fit}.

\addone{The $M$-$R$ fit with two breakpoints is split into three different regimes (segments). These breakpoints correspond to small planets ($M < 4.37 M_\oplus$), intermediate-mass planets ($4.37 M_\oplus < M < 127 M_\oplus$) and giant planets ($ M > 127 M_\oplus$). The first breakpoint has a higher relative uncertainty than the second, with a transition mass of $M_1 = (4.37 \pm 0.72) M_\oplus$. Between the first and second segments, the gradient changes less than the transition from the second to the third segment, making it harder to identify the breakpoint. The fit is particularly suitable for determining both breakpoints and describing the data in the first and second segments. The description of the data in the third segment (the giant planets) is relatively uncertain: These planets show a large scatter, making it difficult to find a well-fitting gradient.}


\begin{table}[h]
    \centering
    \caption{Results for the parameters in Eq. (\ref{eq:piecewise_linear_nbp}) from fitting the exoplanetary mass ($\log(M[M_\oplus])$) and radius ($\log(R[R_\oplus])$) data.}
    \begin{tabular}{|c|c|}
        \hline
        Parameter & Value \\
        \hline
        $c$ & $0.01 \pm 0.01$ \\
        \hline
        $\alpha_1$ & $0.27 \pm 0.04$ \\
        \hline
        $\beta_1$ & $0.40 \pm 0.04$ \\
        \hline
        $\beta_2$ & $-0.72 \pm 0.02$ \\
        \hline
        $\psi_1$ & $0.64 \pm 0.07$ \\
        \hline
        $\psi_2$ & $2.11 \pm 0.03$ \\
        \hline
    \end{tabular}
    \label{tab:results_mass_radius_fit}
\end{table}

The first regime of exoplanets corresponds to planets with masses \addone{below $4.4 M_\oplus$ and roughly follows the relation of $R \propto M^{0.27}$}. These planets are most likely "rocky worlds" with compositions similar to the Earth's. If terrestrial planets can be approximated as constant-density homogeneous spheres, they would follow the simple relation $R \propto M^{1/3}$, which is very similar to what we find. The scatter around this relation in the actual data comes from the differentiated structures of planets and the diversity in their bulk densities, i.e., rocks-to-metals ratios and the possible existence of lighter elements such as water \citep[e.g.,][]{seager_mass-radius_2007,weiss_mass_2013,2016ApJ...819..127Z}.
 
The change in the slope \addone{$\sim4.4 M_\oplus$} defines the transition to the intermediate-mass planets, which could also have non-negligible H-He envelopes. Our fit to the data implies that the maximal mass of "rocky" exoplanets and possibly of naked planetary cores is \addone{somewhere between 4 to $5 M_\oplus$}. This also implies that the minimum mass to accrete a substantial amount of volatile elements is \addone{about 4 to $5 M_\oplus$}. The transition region \addone{is about half of} the theoretical mass limit of about $10 M_\oplus$ for "rocky" exoplanets \citep[e.g.,][]{seager_mass-radius_2007,Fortney2007,2009Natur.462..891C}. The intermediate-mass planets \addone{between $4.4 M_\oplus$ and $127 M_\oplus$} correspond to planets with H-He envelopes but still with a large heavy-element mass fraction. Since the mass range is large, the diversity of the envelope mass fractions varies significantly and can range from very thin atmospheres to rather gaseous envelopes \citep[e.g,][]{weiss_mass_2013,hatzes_definition_2015,2019A&A...630A.135U}. These planets have the steepest $M$-$R$ relation \addone{following $R \propto M^{0.67}$}. An increase in mass results in a significantly larger radius, corresponding mainly to a larger envelope composed of volatile elements. The transition to the gas giants occurs around $127 M_\oplus$ and is where the planets start to be dominated by the H-He envelope. Interestingly, this transition mass is consistent with the suggested transition mass to giant planets based on recent giant planet formation models \citep{2023A&A...675L...8H}.

As expected from their H-He dominated composition, for the giant planets we find that the radius is nearly independent of mass \addone{($R \propto M^{-0.06}$)}. For high-mass objects consisting of a degenerate electron gas, we expect a relation of $R \propto M^{-1/3}$. In the giant planets, the gas is not completely degenerate, leading to a slightly compressible gas and a deviation from the expected relation. At the same time, we also observe a large scatter in radius due to stellar irradiation, different planetary ages, and metallicities which can strongly affect the radii of gas giants \citep[e.g.,][]{2016ApJ...831...64T,2019AJ....158..239T,2020ApJ...903..147M,2023FrASS..1079000M}.

\subsection{The mass-density relation}\label{sec:mass_density_relation}

\begin{figure}[ht]
    \includegraphics[width=\columnwidth]{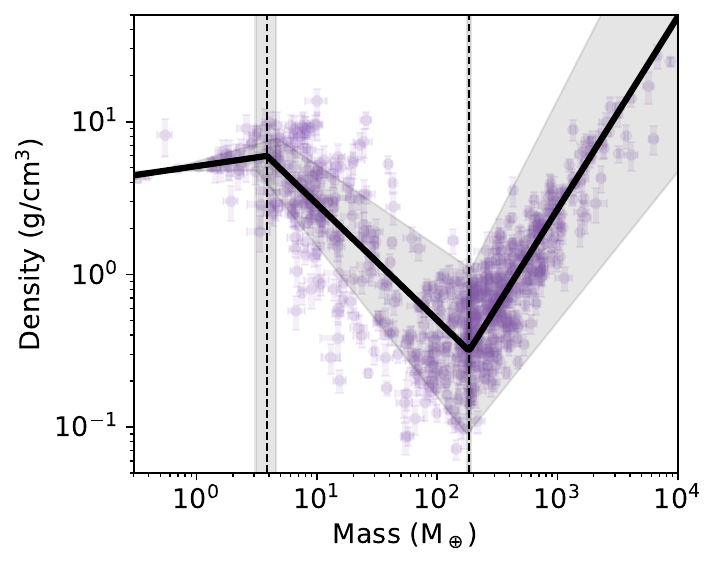}
    \includegraphics[width=\columnwidth]{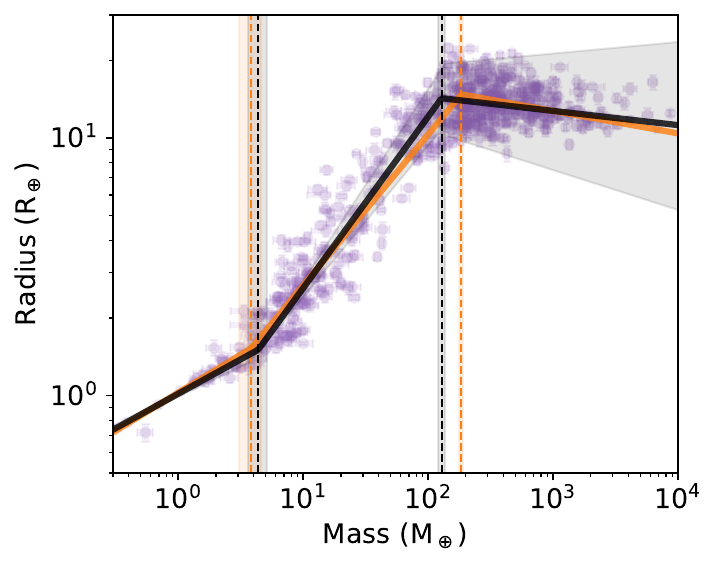}
    \caption{\textbf{Top}: The mass and density data of the exoplanets from the PlanetS Catalog. The mass-density relation is displayed as a solid black line. The dashed lines show the position of the breakpoints. The light-shaded areas are the corresponding $1 \sigma$-uncertainties. \textbf{Bottom}: The mass vs. radius distribution with the best fit for the relation and the breakpoints in black, like in Fig. \ref{fig:mass_radius_fit}. The orange solid and dashed lines show the mass-radius relation and the corresponding breakpoints derived from the fit to the mass-density distribution.}
    \label{fig:mass_density_relation}
\end{figure}

In this subsection, we present a fit to the mass-density ($M$-$\rho$) relation using the planets from the PlanetS Catalog. As in Subsection \ref{sec:mass_radius_relation}, we first determined the best-fit number of breakpoints. Similar to the $M$-$R$ relation, we found that two breakpoints provided the best fit to the data. The fitting function was therefore given by Eq. \ref{eq:piecewise_linear_nbp}), with $\xi = \log(M[M_\oplus])$ and $y = log(\rho[g/cm^3])$ and $n = 2$. The piece-wise linear function with two breakpoints yielded the following power-law $M$-$\rho$ relation:


\begin{equation}
    \rho = 
    \begin{cases}
        (5.12 \pm 0.36) \, M^{(-0.12 \pm 0.11)} & M < (3.84 \pm 0.76) \\
        (16.6 \pm 3.3) \, M^{(-0.76 \pm 0.13)} & (3.84 \pm 0.76) < M < (183 \pm 7) \\
        \frac{(4.39 \pm 2.06)}{10^4} \, M^{(1.26 \pm 0.19)} & M > (183 \pm 7)
    \end{cases}
    \label{eq:mass_density_relation}
\end{equation}

where $\rho$ is in g/cm$^3$ and $M$ in $M_\oplus$. The inferred $M$-$\rho$ fit is shown in the top panel of Fig. \ref{fig:mass_density_relation} together with the data. Tab. \ref{tab:results_mass_density_fit} lists the values of the fitting parameters. \addone{We find that the first breakpoint in the $M$-$\rho$ relation is similar albeit slightly lower than that found in the $M$-$R$ relation. The second breakpoint, the transition to the giant planets, is at a significantly higher mass ($183 M_\oplus$) compared to what we found previously ($127 M_\oplus$)}.

\addone{As discussed above, terrestrial planets can be approximated by a constant density, which is roughly consistent with our findings. The planets in the second and third segments have a large scatter in the $M$-$\rho$ plane, implying that the planetary composition is rather diverse. Consequently, there is also a larger uncertainty in the power-law of the $M$-$\rho$ density for the intermediate and giant planets.}


\begin{table}[ht]
    \centering
    \caption{Results for the parameters in Eq. (\ref{eq:piecewise_linear_nbp}) with $n = 2$ breakpoints from fitting the exoplanetary mass ($\log(M[M_\oplus])$) and density ($\log(\rho[g/cm^3])$) data.}
    \begin{tabular}{|c|c|}
        \hline
        Parameter & Value \\
        \hline
        $c$ & $0.71 \pm 0.03$ \\
        \hline
        $\alpha_1$ & $-0.12 \pm 0.11$ \\
        \hline
        $\beta_1$ & $-0.88 \pm 0.12$ \\
        \hline
        $\beta_2$ & $2.02 \pm 0.05$ \\
        \hline
        $\psi_1$ & $0.59 \pm 0.09$ \\
        \hline
        $\psi_2$ & $2.26 \pm 0.02$ \\
        \hline
    \end{tabular}
    \label{tab:results_mass_density_fit}
\end{table}

To compare the $M$-$\rho$ to the $M$-$R$ relation (as derived in Subsection \ref{sec:mass_radius_relation}), we converted the density $\rho(M)$ from Eq. \ref{eq:mass_density_relation} to a radius using $R = \left(\frac{4 \pi \rho}{3 M}\right)^{1/3}$. Since the $M$-$\rho$ and $M$-$R$ relations were fit with the same data, \addone{they should yield similar results, although some deviations are expected}. \addone{The results are shown} in the bottom panel of Fig. \ref{fig:mass_density_relation}. Indeed, we find that the $M$-$\rho$ and $M$-$R$ relations have very similar behaviors, and the relation derived from the $M$-$\rho$ fit lies well inside the uncertainty of the fit to the $M$-$R$ distribution. The fact \addone{that the power laws were similar while being found separately suggests that our approach yields consistent results. As noted earlier, the largest difference between the two fits is the transition mass between the intermediate and giant planets: Using the $M$-$\rho$ relation yields a significantly higher transition mass.}

\subsection{The radius-density relation}\label{sec:radius_density_relation}

When fitting the $M$-$R$ and the $M$-$\rho$ relations, the three different regimes were defined by a transition mass. \addone{Alternatively,} is also possible to search for transitions in the $R$-$\rho$ relation. Here, we calculated the mean density from the measured $M$ and $R$ (see Section \ref{sec:methods}) and attempted to find a piece-wise linear function that describes the $R$-$\rho$ relation.

\begin{figure}[h]
    \includegraphics[width=\columnwidth]{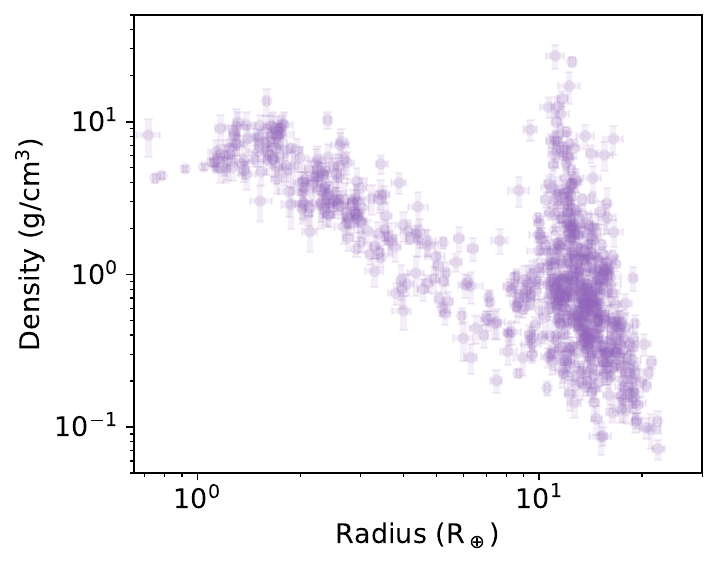}
    \includegraphics[width=\columnwidth]{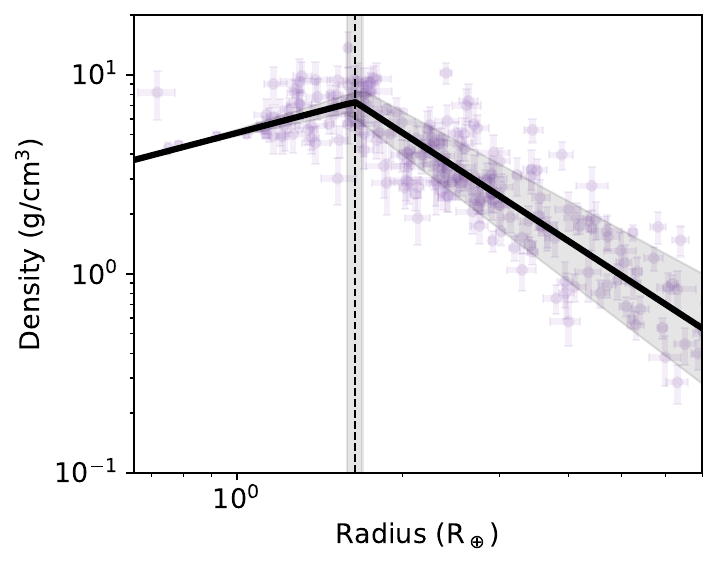}
    \caption{\textbf{Top}: Radius-density distribution of the exoplanets from the PlanetS Catalog. \textbf{Bottom}: The exoplanets from the PlanetS Catalog smaller than $7 R_\oplus$ are displayed together with the mass-density relation (solid black line) and the corresponding breakpoints (dashed line). The light-shaded areas are the $1 \sigma$-uncertainties.}
    \label{fig:radius_density_relation}
\end{figure}

The data are shown in the top panel of Fig. \ref{fig:radius_density_relation}. Qualitatively, three regimes can be identified: For the smallest planets, the density seems nearly independent of radius. Then, there is a breakpoint where the density decreases steeply with increasing radius. The giant planets (around $13 R_\oplus$) show a \addone{strong} dispersion in density. This is similar to what we already observed in the $M$-$R$ relation: For the giant planets, the bulk density can vary greatly due to their age, instellation flux, and \addone{composition}. The intermediate and the giant planets start to overlap around $7 R_\oplus$ \addone{and the density appears almost uncorrelated with the radius.}. Therefore, for fitting the $R$-$\rho$ relation, we excluded planets larger than $7 R_\oplus$.

\addone{We found that the best model (with the lowest BIC) to describe the $R$-$\rho$ relation uses one breakpoint.} This is unlike the two breakpoints for the $M$-$R$ and $M$-$\rho$ relations. However, it is somewhat expected since we excluded the giant planets. The resulting $R$-$\rho$ relation is:


\begin{equation}
    \rho =
    \begin{cases}
        (5.11 \pm 0.19) \, R^{(0.73 \pm 0.15)} & R < (1.64 \pm 0.05) \\
        (17.9 \pm 1.5) \, R^{(-1.80 \pm 0.17)} & R > (1.64 \pm 0.05)
    \end{cases}
    \label{eq:radius_density_relation}
\end{equation}

where $\rho$ and $R$ are in g/cm$^3$ and $R_\oplus$, respectively. The $R$-$\rho$ best-fit and the data are shown in the bottom panel of Fig. \ref{fig:radius_density_relation}.

The values of the parameters for the piece-wise linear function with one breakpoint (see Eq. \ref{eq:piecewise_linear_1bp}) are listed in Tab. \ref{tab:results_radius_density_fit}. 


\begin{table}
    \centering
    \caption{Results for the parameters in Eq. (\ref{eq:piecewise_linear_1bp}) from fitting the exoplanetary radius ($\log(R[R_\oplus])$) and density ($\log(\rho[g/cm^3])$) data.}
    \begin{tabular}{|c|c|}
        \hline
        Parameter & Value \\
        \hline
        $c$ &  $ 0.71 \pm 0.02$ \\
        \hline
        $\alpha_1$ & $0.73 \pm 0.15$ \\
        \hline
        $\beta_1$ & $-2.53 \pm 0.16$ \\
        \hline
        $\psi_1$ & $0.22 \pm 0.01$ \\
        \hline
    \end{tabular}
    \label{tab:results_radius_density_fit}
\end{table}

\addone{A notable} result is the \addone{breakpoint at around $1.6 R_\oplus$. Its relative uncertainty of 3\% is significantly lower than for the mass threshold between the small and intermediate planets derived from the $M$-$R$ relation (14\%) or the $M$-$\rho$ relation (20\%)}. This shows that it is beneficial to consider the radius when distinguishing between different planetary types \citep{rogers_most_2015,lozovsky_threshold_2018}. Similar to the results for the $M$-$\rho$ relation (see Subsection \ref{sec:mass_density_relation}), \addone{the relative uncertainty of the $R$-$\rho$ in the first segment is rather high, but it is consistent with a constant density approximation.} 

\citet{lozovsky_threshold_2018} found threshold radii above which a certain composition is unlikely. For purely rocky planets they found a threshold radius of $1.66^{+0.01}_{-0.08} R_\oplus$. Larger planets must consist at least partly of lighter elements, such as H and He. This is consistent with our result of a breakpoint \addone{at $1.64 R_\oplus$}. However, they only distinguished between super-Earths and mini-Neptunes at $\simeq 3R_\oplus$, because planets with a larger radius have a substantial H-He atmosphere (at least $2\%$ mass fraction). In contrast, based on our data no distinction can be made there.

Our result of the radius breakpoint \addone{at $1.64 R_\oplus$} also coincides with the position of the radius valley around $1.5 - 2 R_\oplus$. The radius valley is a bimodal feature in the occurrence rate of planets as a function of their radii, \addone{which manifests itself as a} scarcity of planets with $R \simeq 1.5 - 2 R_\oplus$. It has been observed for planets with short periods \citep[e.g.,][]{fulton_california-kepler_2017} and is often used for the distinction between super-Earths (below the valley) and mini-Neptunes (above the valley).

Several previous studies have shown how photoevaporation or core-powered mass loss can lead to the depletion of the gaseous envelopes of planets at such radii \citep{2016ApJ...831..180C,owen_evaporation_2017,2020A&A...643L...1V}, and therefore explain the radius valley. In particular, \citet{2022A&A...668A.178K} suggested that the $M$-$R$ relation of intermediate planets is shaped by their thermal evolution and hydrodynamic escape. Additionally, it has also been suggested that the planets at the upper edge of the radius valley are helium-rich \citep{2023NatAs...7...57M}. This suggests that due to evaporation of the gaseous envelope for masses with $R \simeq 1.3 R_\oplus$ planets are naked rocky cores, while around $2.6 R_\oplus$ they sustain at least part of their H-He envelope. As an alternative, it has also been suggested that the bimodal radius distribution of planets smaller than about $4 R_\oplus$ is due to different compositions of rocky super-Earths and ice- or water-rich mini-Neptunes \citep{2019PNAS..116.9723Z,2020A&A...643L...1V,2021A&A...650A.152I,2022ApJ...939L..19I}.

\subsubsection{Comparison with previous studies}\label{sec:comparison}

A comparison of our results with previous studies is presented in Tab. \ref{tab:compare_results}. To facilitate the comparison, the relations and breakpoints were converted to Earth units ($M_\oplus$ and $R_\oplus$). For the $M$-$R$ relation, we use the results from Subsection \ref{sec:mass_radius_relation}. Also, the mass-density relation from \citet{hatzes_definition_2015} was converted to a $M$-$R$ relation. From \citet{edmondson_breaking_2023} we chose the $M$-$R$ relation for the giant planets instead of their mass-radius-temperature relation.

\begin{figure}[h]
    \includegraphics[width=\columnwidth]{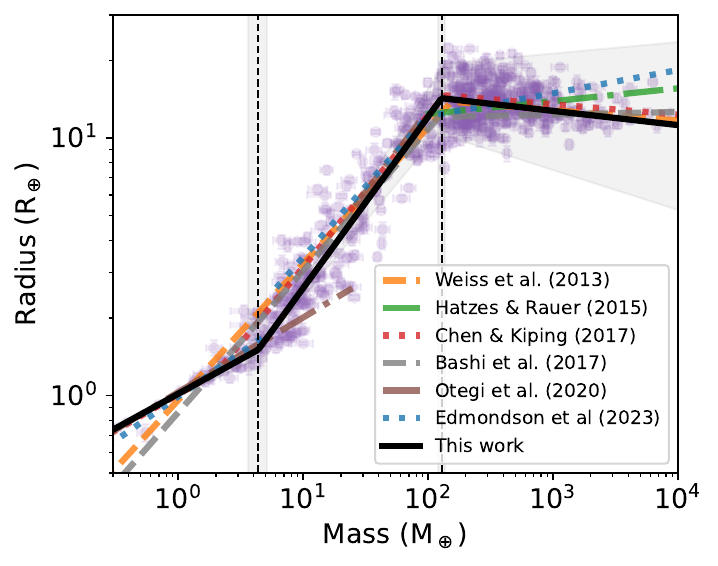}
    \caption{Comparison of the $M$-$R$ relations from different studies with the exoplanetary data from the PlanetS Catalog.}
    \label{fig:compare_results}
\end{figure}

Overall, it is clear that there is a rather good agreement between the various studies despite the use of different methods and data. The $M$-$R$ relations from the different studies are shown in Fig. \ref{fig:compare_results} together with the data from the PlanetS Catalog. It can be seen that the relations from \citet{weiss_mass_2013} and \citet{bashi_two_2017} underestimate the radii of the smallest planets and are not a good fit. This is because they only use one breakpoint in the $M$-$R$ relation, \addone{corresponding} to the transition to giant planets. The relation by \citet{chen_probabilistic_2016} does not fit the planets around $10M_\oplus$ very well, because the location of the transition from small to intermediate-mass planets is underestimated. \citet{hatzes_definition_2015} do not fit planets below $95 M_\oplus$ at all. The relation by \citet{otegi_revisited_2020} remains a good fit for the data set. The main difference to our results is the transition mass from small to intermediate planets. \citet{otegi_revisited_2020} defined the transition with the water-composition line, while we used a statistical approach to find this transition. The benefit of our approach is that it does not rely on \addone{or a priori assumptions} or theoretical models to \addone{determine} the transition (and the associated uncertainties in, e.g., the EOS). \addone{Perhaps surprisingly, despite the} spread of the planets in the $M$-$R$ diagram, our data set yielded a \addone{rather small uncertainty in the transition masses}. Similarly, \citet{edmondson_breaking_2023} used a pure-ice EOS to mark the transition between small and intermediate planets, which leads to a good description of the smallest planets and intermediate planets. However, between $\sim 30 - 90 M_\oplus$ their relation for icy planets significantly under-predicts the radii of the planets in the PlanetS Catalog, leading to a poor fit. Compared to all the listed previous small-planet $M$-$R$ fits, our uncertainty on the power-law index is \addone{slightly} higher. This is likely \addone{because the small planets in our updated data have diverse radii.} For the giant exoplanets, all the relations can qualitatively describe the $M$-$R$ relation, although they are quite different and can even have a different sign of the gradient. The data in this regime show a \addone{strong} dispersion, which leads to rather large uncertainties in the fitted relation. \addone{Interestingly, using the $M$-$\rho$ relation to determine the transition to the giant planets yielded a significantly larger mass ($183 M_\oplus$) than both the $M$-$R$ fit ($127 M_\oplus$) and results from previous studies ($95$ to $150 M_\oplus$).}

\section{Discussion and conclusions}\label{sec:discussion}

\addone{In this paper, we use the updated PlanetS Catalog to infer $M$-$R$, \addone{$M$-$\rho$, and $R$-$\rho$} relations and determined the transitions between different planetary types.} While the presented analysis provides insight into the different planet regimes, it was simplified and did not consider all the subtleties related to exoplanetary data. First, we treated the data as one unit although it is clear that the data set is inhomogeneous and combines different observational methods with different biases. The effects of observational bias for the most part have not been considered. 

Other parameters affect the $M$-$R$ relation that were not investigated in this work. For example, for giant planets stellar age and irradiation are important. Giant planets are massive enough that their self-gravity causes them to contract over long timescales ($\sim 1$ Gyr; \citet[e.g.,][]{1977Icar...30..305H,2001RvMP...73..719B}), and therefore their radius is expected to be correlated with their age. Additionally, high instellation fluxes are inflating the radii of warm giant planets \citep[e.g.,][]{Guillot2006,Fortney2007,Fortney2010,2016ApJ...831...64T,2023FrASS..1079000M}. This effect was included in \citet{weiss_mass_2013} and \citet{edmondson_breaking_2023}, where a third parameter (instellation flux or equilibrium temperature) was added to better fit the $M$-$R$ of the giant planets. Recently, there have been also attempts to move beyond the two-parametric $M$-$R$ relationship: For example, \citet{2023ApJ...956...76K} presented a framework to characterize exoplanets using up to four simultaneous parameters. In the future, such approaches may better constrain the transition from small- to intermediate- planets from observational data by considering additional parameters. 

Out of the over 5000 detected exoplanets, only 688 have robust enough mass and radius measurements to be included in the PlanetS Catalog. While this means that only a fraction of the currently detected exoplanets were used in this work, the results are also more robust, since only planets with low mass and radius uncertainties are included. More accurate data are needed to analyze the whole parameter space occupied by exoplanets.

The key results from our study can be summarized as follows:

\begin{enumerate}
    \item 
    \addone{Our analysis yielded} a small-to-intermediate transition mass \addone{of $(4.37 \pm 0.72) M_\oplus$. Small planets below the transition mass follow $R \propto M^{0.27}$}. These are "rocky worlds" with different bulk compositions. \addone{The transition to the intermediate-mass could imply a maximal mass of $\sim4.4 M_\oplus$ of "rocky" exoplanets and naked planetary cores}. 
    \item The transition from rocky to volatile-rich planets can also be defined in terms of the radius. By fitting the radius-density relation, we found that the transition occurs around $1.64 \pm 0.05 R_\oplus$. The transition in radius has \addone{a significantly lower relative uncertainty} than the one in mass. Furthermore, the transition radius is consistent with the radius valley around $1.5$ to $2 M_\oplus$. 
    \item Intermediate-mass planets ranging from about $4.4 M_\oplus$ to $127 M_\oplus$ behave as $R \propto M^{0.67}$. They correspond to planets with H-He envelopes. The transition to giant planets occurs at $(127 \pm 7) M_\oplus$ and corresponds to planets that are H-He-rich.
    \item \addone{Using the $M$-$\rho$ relation to find the transition to the giant planets yielded a significantly higher transition mass of $(183 \pm 7) M_\oplus$.} 
    \item The radii of giant planets are nearly independent of their masses and the mass-radius relation in this regime follows $R \propto M^{-0.06}$.
    \item Overall, planets of different compositions and structures can have the same mass and radius. This leads to an intrinsic degeneracy of the mass-radius distribution of exoplanets.
    \end{enumerate}

\clearpage
Ongoing and future observations on the ground and in space will improve our understanding of exoplanets. The James Webb Space Telescope \citep{Gardner2006} and the Ariel mission \citep{Tinetti2018} will enable us to characterize the atmospheres of transiting planets, providing information about their chemical compositions. High-resolution spectroscopy from current (e.g., SPIROU \citet{2014SPIE.9147E..15A}, CARMENES \citet{2016SPIE.9908E..12Q}) and future (e.g., NIRPS \citet{2017Msngr.169...21B,2017SPIE10400E..18W}, CRIRES+ \citet{2004SPIE.5492.1218K,2014Msngr.156....7D,2023arXiv230108048D}) ground-based telescopes will provide further improvements with accurate radial-velocity measurements and atmospheric characterizations. 

More exoplanets detected via direct imaging e.g. by SPHERE at the Very Large Telescope \citep{beuzit_sphere_2019} will facilitate studying the properties of planets on wide orbits. Also, the upcoming PLATO mission \citep{2014ExA....38..249R} will detect and characterize small terrestrial planets as well as intermediate-mass and giant planets. Theoretical studies to understand the key physical processes that shape the exoplanetary populations are also being developed, and we hope to be able to connect the properties of exoplanets with their origin and evolution. These ongoing and upcoming efforts are expected to reveal new insights into the population of planets beyond the solar system.

\begin{acknowledgements}
    \addone{We thank the anonymous reviewer for useful feedback.} We acknowledge support from SNSF grant \texttt{\detokenize{200020_188460}} and the National Centre for Competence in Research ‘PlanetS’ supported by SNSF. This research used data from the NASA Exoplanet Archive, which is operated by the California Institute of Technology, under contract with the National Aeronautics and Space Administration under the Exoplanet Exploration Program. 
    Extensive use was also made of the Python packages \textit{NumPy} \citep{harris2020array}, SciPy \citep{2020SciPy-NMeth}, Jupyter \citep{jupyter}, \textit{Matplotlib} \citep{Hunter2007}, \textit{pandas} \citep{reback2020pandas}, and \textit{piecewise-regression} \citep{pilgrim_piecewise-regression_2021}.
\end{acknowledgements}

\begin{table*}
    \caption{Results for the mass-radius relations and corresponding breakpoints (transitions) from different studies and this work. The masses and radii are in Earth units.}
    \small
    \renewcommand{\arraystretch}{1.3}
    \begin{tabular}{|c|c|c|c|c|c|c|}
    \hline
    & \multicolumn{2}{|c|}{Small planets} & \multicolumn{2}{|c|}{Intermediate planets} & \multicolumn{2}{|c|}{Giant planets} \\ \hline
    Source & \multicolumn{2}{|c|}{Mass-radius : $R(M)$} & Transition & Mass-radius: $R(M)$& Transition & Mass-radius: $R(M)$ \\ \hline
    \citet{weiss_mass_2013} & \multicolumn{2}{|c|}{$ - $} & $ - $ & $0.96^{+0.08}_{-0.07} M^{0.53 \pm 0.05}$ & $M = 150^\dag$ & $16.9^{+4.5}_{-3.6} M^{-0.04 \pm 0.01}$ \\ \hline
    \citet{hatzes_definition_2015} & \multicolumn{2}{|c|}{$ - $} & $ - $ & $ - $ & $M = 95.3^\dag$ & $9.83^{+0.34}_{-0.35} M^{0.05 \pm 0.01}$ \\ \hline
    \citet{chen_probabilistic_2016} & \multicolumn{2}{|c|}{$1.01 \pm 0.05 M^{0.28 \pm 0.01}$} & $M = 2.04^{+0.66}_{-0.59}$ & $0.81 \pm 0.05 M^{0.59^{+0.04}_{-0.03}}$ & $M = 132^{+18}_{-21}$ & $17.8^{+9.7}_{-5.9} M^{-0.04 \pm 0.02}$ \\ \hline
    \citet{bashi_two_2017} & \multicolumn{2}{|c|}{$ - $} & $ - $ & $0.86^{+0.08}_{-0.07} M^{0.55 \pm 0.02}$ & \makecell{$M = 124 \pm 7$ \\ $R = 12.1 \pm 0.5$} & $11.5 \pm 0.6 M^{0.01 \pm 0.02}$ \\ \hline
    \citet{otegi_revisited_2020} & \multicolumn{2}{|c|}{$1.03 \pm 0.02 M^{0.29 \pm 0.01}$} & Water line & $0.70 \pm 0.11 M^{0.63 \pm 0.04}$ & $ - $ & $ - $ \\ \hline
    \citet{edmondson_breaking_2023} & \multicolumn{2}{|c|}{$0.99 \pm 0.02 M^{0.34 \pm 0.01}$} & Pure-ice EOS & $0.97 \pm 0.07 M^{0.55 \pm 0.02}$ & $M = 115 \pm 19$ & $8.01 \pm 0.48 M^{0.09 \pm 0.001}$ \\ \hline
    This work & \multicolumn{2}{|c|}{$1.02 \pm 0.03 M^{0.27 \pm 0.04}$} & \makecell{$M = 4.37 \pm 0.72$ \\ $R = 1.64 \pm 0.05$} & $0.56 \pm 0.03 M^{0.67 \pm 0.05}$ & $M = 127 \pm 7$ & $18.6 \pm 6.7 M^{-0.06 \pm 0.07}$ \\ \hline 
    \multicolumn{2}{l}{$^{\dag}$\footnotesize{Note that no uncertainties on these transition points were provided.}}
    \end{tabular}
    \label{tab:compare_results}
 \end{table*}

\bibliographystyle{aa}
\bibliography{ref} 

\end{document}